# Analytic solution of the Schrödinger equation for an electron in the field of a molecule with an electric dipole moment

A. D. Alhaidari

*Shura Council, Riyadh 11212, Saudi Arabia*
AND
*Physics Department, King Fahd University of Petroleum & Minerals, Dhahran 31261, Saudi Arabia*

email: haidari@mailaps.org

We relax the usual diagonal constraint on the matrix representation of the eigenvalue wave equation by allowing it to be tridiagonal. This results in a larger solution space that incorporates an exact analytic solution for the non-central electric dipole potential $\cos\theta/r^2$, which was known not to belong to the class of exactly solvable potentials. As a result, we were able to obtain an exact analytic solution of the three-dimensional time-independent Schrödinger equation for a charged particle in the field of a point electric dipole that could carry a nonzero net charge. This problem models the interaction of an electron with a molecule (neutral or ionized) that has a permanent electric dipole moment. The solution is written as a series of square integrable functions that support a tridiagonal matrix representation for the angular and radial components of the wave operator. Moreover, this solution is for all energies, the discrete (for bound states) as well as the continuous (for scattering states). The expansion coefficients of the radial and angular components of the wavefunction are written in terms of orthogonal polynomials satisfying three-term recursion relations. For the Coulomb-free case, where the molecule is neutral, we calculate critical values for its dipole moment below which no electron capture is allowed. These critical values are obtained not only for the ground state, where it agrees with already known results, but also for excited states as well.

## I. INTRODUCTION

The interaction of a charged particle with an electric dipole is a fundamental problem, which received a lot of attention since the early days of nuclear and molecular physics [1]. In the latter, this interest was fueled by the observation that electron capture by a frozen molecule with a permanent electric dipole moment could take place only if the dipole moment exceeds a certain minimum critical value [2]. Another source of interest came from early experimental work on the scattering of low energy electrons by such molecules, which exhibited peculiar features [3]. Moreover, the electron binding properties of polar molecules have been the topic of considerable theoretical and experimental interest [4]. It has been shown that the electric dipole field is capable of supporting an infinite number of bound states for an electron if the dipole moment is greater than the critical value 0.6393 a.u. (or 1.625D, where 1D = $10^{-18}$ esu-cm). Under these circumstances, the excess electron will be bound to the molecular dipole field, giving rise to the so called dipole-bound anion. The critical dipole moment (its minimum value below which no bound states could be supported) does not depend on the size of the dipole [2]. However, if the system is treated dynamically to include the rotational degrees of freedom of the nuclei, the infinite number of bound states for the charged electron in the dipolar molecular field reduces to a finite number [5]. From the experimental point of view, the formation of a dipole bound anion has been difficult to observe because the much diffused and loosely bound electrons are easily stripped away by thermal collisions and/or by the electric fields to which they are exposed. However, recent experimental advances have made it possible to measure the energy of one dipole-bound electron very accurately [4].

Despite renewed interest in this problem, no exact analytic solution (aside from our recent contribution [6]) was reported in the literature. This is because, even in the ideal case of a point electric dipole, the interaction includes the intractable noncentral potential $\cos\theta/r^2$ (in spherical coordinates) which was known not to belong to any of the established classes of exactly solvable potentials. In this paper, we give the details of our investigation and findings in [6]. As a result of this investigation, we did confirm and explain the outstanding difficulty that the energy eigenvalue equation $H|\phi_n\rangle = E_n|\phi_n\rangle$ for this problem does not have a closed form solution for non-zero dipole moment. However, our approach, follows another path in which we employ the tools of our "Tridiagonal Physics" program [7]. In this program, we relax the restriction of a diagonal representation on the solution space of the eigenvalue wave equation. We only require that the matrix representation of the wave operator (both components; the radial and angular) be tridiagonal and symmetric. That is, the action of the wave operator on the elements of the basis is not restricted to be of the form $(H-E)|\phi_n\rangle \sim |\phi_n\rangle$ but allowed to take the less constrained form $(H-E)|\phi_n\rangle \sim |\phi_n\rangle + |\phi_{n-1}\rangle + |\phi_{n+1}\rangle$. Therefore, in a suitable orthogonal basis, this gives the following elements of the tridiagonal matrix representation of the wave operator

$$\langle\phi_n|H-E|\phi_m\rangle \sim (a_n - E)\delta_{n,m} + b_n\delta_{n,m-1} + b_{n-1}\delta_{n,m+1}, \qquad (1.1)$$

where the coefficients $\{a_n, b_n\}_{n=0}^{\infty}$ are real and, in general, functions of the energy, angular momentum, and potential parameters. Thus, the matrix wave equation, which is obtained by expanding the wavefunction $|\psi\rangle$ as $\sum_m f_m|\phi_m\rangle$ in $(H-E)|\psi\rangle = 0$ and projecting on the left by $\langle\phi_n|$, results in the following three-term recursion relation

$$E f_n = a_n f_n + b_{n-1} f_{n-1} + b_n f_{n+1}. \qquad (1.2)$$

Consequently, the problem translates into finding solutions of this symmetric three-term recursion relation for the expansion coefficients of the wavefunction. The original contribution of our work is that by relaxing the usual constraint on the matrix representation of the wave operator, from being diagonal to allow for a tridiagonal representation, the solution space becomes large enough to incorporate an exact analytic solution for the noncentral potential $V(r,\theta) = Z/r + \xi\cos\theta/r^2$, where $\xi$ is the electric dipole moment and $Z$ is its net charge. Now, it is obvious that the solution of (1.2) is obtained modulo an overall non-singular factor, which is a function of the physical parameters of the problem, but otherwise independent of $n$. Uniqueness of the solution is achieved, for example, by the requirement of normalizability of the wavefunction or by making the typical seed assignment $f_0 = 1$. The latter choice makes $f_n$ a polynomial of degree $n$ in $E$. Moreover, one may note that the solution of the problem as depicted by Eq. (1.2) is obtained for all $E$, the discrete as well as the continuous. Additionally, the representation (1.1) clearly shows that the discrete spectrum is easily obtained by the diagonalization constraint, which requires that

$b_n = 0$, and $a_n - E = 0$. (1.3)

for all $n$.

In Sec. III, we demonstrate clearly that by relaxing the diagonal constraint on the matrix representation of the angular component of the wave operator the solution space becomes larger incorporating the noncentral dipole potential $\cos\theta/r^2$. The associated orthogonal polynomials (expansion coefficients of the angular wavefunction) that satisfy the resulting three-term recursion relation are given in terms of the dipole moment and the azimuthal quantum number. They are not classical polynomials but are completely specified. We refer to them as the "*dipole polynomials*". Thereafter, we obtain in Sec. IV the radial component of the wave function for the continuum as well as the bound states. The pure dipole interaction (for the case of a neutral molecule) is studied in Sec. V where we obtain critical values for the dipole moment below which no electron capture could take place. These values are obtained not only for the ground state, where it agrees with already known results, but also for excited states as well. We end the paper with summary and concluding remarks in Sec. VI.

## II. PRELIMINARIES

The three-dimensional time-independent Schrödinger equation for a particle of mass $M$ and charge $q$ in an electrostatic potential $V(\vec{r})$ is

$$\left[ -\tfrac{\hbar^2}{2M}\vec{\nabla}^2 + qV(\vec{r}) - E \right]|\psi\rangle = 0, \quad (2.1)$$

where $\vec{\nabla}$ is the three-dimensional Laplacian and the energy, $E$, is either discrete or continuous. In the spherical coordinates $\vec{r} = \{r, \theta, \phi\}$, this wave equation could be written explicitly as follows

$$\left\{ \frac{1}{r^2}\frac{\partial}{\partial r}r^2\frac{\partial}{\partial r} + \frac{1}{r^2}\left[ (1-x^2)\frac{\partial^2}{\partial x^2} - 2x\frac{\partial}{\partial x} + \frac{1}{1-x^2}\frac{\partial^2}{\partial \phi^2}\right] - 2V + 2E\right\}|\psi\rangle = 0, \quad (2.2)$$

where $x = \cos\theta$ and we have used the atomic units $\hbar = M = q = 1$. Consequently, this differential equation is separable for potentials of the form

$$V(\vec{r}) = V_r(r) + \frac{1}{r^2}\left[V_\theta(x) + \frac{1}{1-x^2}V_\phi(\phi)\right]. \quad (2.3)$$

This is so because if we write the wavefunction as $\psi(r,\theta,\phi) = r^{-1}R(r)\Theta(\theta)\Phi(\phi)$, then the wave equation (2.2) with the potential (2.3) becomes separated in all three coordinates as follows

$$\left(\frac{d^2}{d\phi^2} - 2V_\phi + 2E_\phi\right)|\Phi\rangle = 0, \quad (2.4a)$$

$$\left[(1-x^2)\frac{d^2}{dx^2} - 2x\frac{d}{dx} - \frac{2E_\phi}{1-x^2} - 2V_\theta + 2E_\theta\right]|\Theta\rangle = 0, \quad (2.4b)$$

$$\left(\frac{d^2}{dr^2} - \frac{2E_\theta}{r^2} - 2V_r + 2E\right)|R\rangle = 0, \quad (2.4c)$$

where $E_\phi$ and $E_\theta$ are the separation constants, which are real and dimensionless. Square integrability is with respect to the following integration measures

$$\int |\psi|^2 \, d^3\vec{r} = \int_0^\infty |R|^2 \, dr \int_{-1}^{+1} |\Theta|^2 \, dx \int_0^{2\pi} |\Phi|^2 \, d\phi. \tag{2.5}$$

The components of the wavefunction are also required to satisfy the boundary conditions: $R(0) = R(\infty) = 0$, $\Phi(\phi) = \Phi(\phi + 2\pi)$, $\Theta(0)$ and $\Theta(\pi)$ are finite. Now we specialize to the case where the charged particle moves under the influence of a point electric dipole of moment $p$ pointed along the positive $z$-axis, fixed at the origin, and carries a net charge $Zq$. As such, the problem could be used to model the interaction of an electron with ionized molecule that has a permanent electric dipole moment. Therefore, the potential components in (2.3) read as follows: $V_\phi = 0$, $V_\theta = (p/qa_0)\cos\theta$, and $V_r = Z/a_0 r$, where $a_0 = 4\pi\epsilon_0 \hbar^2 / Mq^2$ (for an electron it is the Bohr radius). Measuring length (e.g., $r$ and $p/q$) in units of $a_0$ then the energy $E$ will be given in units of $(\hbar/a_0)^2 M^{-1}$ and $V_\theta \to (p/q)\cos\theta$, $V_r \to Z/r$. Moreover, the set of equations (2.4a-2.4c) show that $E_\theta$ is a function of the electric dipole moment $p$ and $E_\phi$ and that the system's energy $E$ is a function of the physical parameters as $E(E_\theta, Z) = E(E_\phi, p, Z)$. In the following two sections, we obtain the angular and radial components of the wavefunction. These will be written as series of discrete square integrable functions that support a tridiagonal matrix representation for the corresponding wave operator.

### III. ANGULAR COMPONENT OF THE WAVEFUNCTION

It is easy and straightforward to obtain the normalized solution of Eq. (2.4a) with $V_\phi = 0$ that satisfies the boundary conditions as

$$\Phi_m(\phi) = \frac{1}{\sqrt{2\pi}} e^{\pm im\phi}, \quad m = 0, 1, 2, \ldots, \tag{3.1}$$

giving $E_\phi = \frac{1}{2} m^2$. Now, we expand the angular component of the wavefunction, $\Theta(\theta)$, in a complete square integrable basis functions $\{\chi_n(x)\}_{n=0}^\infty$ as $\Theta(\theta) = \sum_{n=0}^\infty f_n^m(E_\theta) \chi_n(x)$. These $L^2$ basis elements must satisfy the boundary conditions in the configuration space with coordinate $x \in [-1, +1]$. They are written as

$$\chi_n(x) = a_n (1-x)^\alpha (1+x)^\beta P_n^{(\mu,\nu)}(x), \tag{3.2}$$

where $P_n^{(\mu,\nu)}(x)$ is the Jacobi polynomial of degree $n = 0, 1, 2, \ldots$ The dimensionless real parameters $\alpha, \beta \geq 0$, $\mu, \nu > -1$ and $a_n$ is the normalization constant associated with the orthogonality of the Jacobi polynomials,

$$a_n = \sqrt{\frac{2n+\mu+\nu+1}{2^{\mu+\nu+1}} \frac{\Gamma(n+1)\Gamma(n+\mu+\nu+1)}{\Gamma(n+\mu+1)\Gamma(n+\nu+1)}}. \tag{3.3}$$

Using the differential equation (A.3) and differential formula (A.4) for the Jacobi polynomials, shown in Appendix A, we obtain

$$\left[ (1-x^2) \frac{d^2}{dx^2} - 2x \frac{d}{dx} \right] |\chi_n\rangle = \left[ -n \left( x + \frac{\nu-\mu}{2n+\mu+\nu} \right) \left( \frac{\mu-2\alpha}{1-x} + \frac{2\beta-\nu}{1+x} \right) + \alpha^2 \frac{1+x}{1-x} + \beta^2 \frac{1-x}{1+x} \right.$$
$$\left. - (2\alpha\beta + \alpha + \beta) - n(n+\mu+\nu+1) \right] |\chi_n\rangle + 2 \frac{(n+\mu)(n+\nu)}{2n+\mu+\nu} \left( \frac{\mu-2\alpha}{1-x} + \frac{2\beta-\nu}{1+x} \right) \frac{a_n}{a_{n-1}} |\chi_{n-1}\rangle \tag{3.4}$$

Therefore, the action of the differential operator of Eq. (2.4b) on the basis element (3.2) reads as follows

$$(H_\theta - E_\theta)|\chi_n\rangle = \left[\frac{n}{2}\left(x + \frac{\nu-\mu}{2n+\mu+\nu}\right)\left(\frac{\mu-2\alpha}{1-x} + \frac{2\beta-\nu}{1+x}\right) - \frac{\alpha^2}{2}\frac{1+x}{1-x} - \frac{\beta^2}{2}\frac{1-x}{1+x} + \frac{E_\phi}{1-x^2} + V_\theta \right.$$
$$\left. -E_\theta + \alpha\beta + \frac{\alpha+\beta}{2} + \frac{n}{2}(n+\mu+\nu+1)\right]|\chi_n\rangle - \frac{(n+\mu)(n+\nu)}{2n+\mu+\nu}\left(\frac{\mu-2\alpha}{1-x} + \frac{2\beta-\nu}{1+x}\right)\frac{a_n}{a_{n-1}}|\chi_{n-1}\rangle \quad (3.5)$$

The recurrence relation (A.1) and orthogonality formula (A.5) of the Jacobi polynomials show that a tridiagonal matrix representation of the angular component of the wave operator, $\langle\chi_n|H_\theta - E_\theta|\chi_{n'}\rangle$, is achievable with $V_\theta \sim x$ if and only if $\mu = \nu = m$ and $\alpha = \beta = \frac{1}{2}m$. To obtain an explicit expression for the matrix elements $\langle\chi_n|H_\theta - E_\theta|\chi_{n'}\rangle$, we project the action of the angular wave operator given by (3.5) on the basis elements from left. Employing again the orthogonality property and recurrence relation of the Jacobi polynomials, we obtain the following tridiagonal structure of the wave operator

$$2\langle\chi_n|H_\theta - E_\theta|\chi_{n'}\rangle = \left[\left(n+m+\tfrac{1}{2}\right)^2 - \left(\gamma+\tfrac{1}{2}\right)^2\right]\delta_{n,n'}$$
$$-\xi\sqrt{\frac{n(n+2m)}{(n+m)^2-1/4}}\delta_{n,n'+1} - \xi\sqrt{\frac{(n+1)(n+2m+1)}{(n+m+1)^2-1/4}}\delta_{n,n'-1} \quad (3.6)$$

where $\xi = p/q$ and we have introduced the dimensionless real parameter $\gamma$ by writing $2E_\theta \equiv \gamma(\gamma+1)$. Equation (2.4c) shows that $\gamma$ plays the role of the angular momentum quantum number $\ell$ in spherically symmetric problems. However, unlike $\ell$ that assumes non-negative integral values, $\gamma$ is a continuous parameter that could be positive or negative. For positive parameter $E_\theta$, $\gamma$ must be either greater than zero or less than $-1$. Another argument in support of this range of values of $\gamma$ goes as follows. For an arbitrary dipole moment $\xi$, Eq. (3.6) shows that as the integers $n$ and $m$ increase the representation changes its signature (becoming indefinite) when crossing the threshold defined by

$$\left(n+m+\tfrac{1}{2}\right)^2 \leq \left(\gamma+\tfrac{1}{2}\right)^2. \quad (3.7)$$

Consequently, to keep the representation, which is bounded from below, definite and prevent it from degenerating the real parameter $\gamma$ must satisfy either $\gamma \geq 0$ or $\gamma \leq -1$. However, we will also show in Sec. V (when discussing the special case of pure dipole interaction) that $\gamma = -\frac{1}{2}$ is a singular case that corresponds to critical values of the dipole moment below which no bound states could be formed [2].

The tridiagonal matrix representation of the angular wave operator in (3.6) makes the wave equation (2.4b) equivalent to the following three-term recursion relation for the expansion coefficients of the angular component of the wavefunction

$$\left(\gamma+\tfrac{1}{2}\right)^2 f_n^m = \left(n+m+\tfrac{1}{2}\right)^2 f_n^m - \xi\sqrt{\frac{n(n+2m)}{(n+m)^2-1/4}}f_{n-1}^m - \xi\sqrt{\frac{(n+1)(n+2m+1)}{(n+m+1)^2-1/4}}f_{n+1}^m. \quad (3.8)$$

This relation implies that if $\xi$ becomes too large then reality of the representation will be violated. To maintain reality, the dipole moment should not exceed a certain critical value. For a fixed integer $m$ and for $n = 0,1,2,...$, this critical value is the smallest $\xi$ that makes the right side of (3.8) vanish. It depends on the azimuthal quantum number $m$ and we denote it by $\hat{\xi}_m$. In Sec. V, we show how to calculate these critical values. Table 1 is a list of some of these values in atomic units calculated to the desired accuracy. Rewriting the expansion coefficients $f_n^m(E_\theta)$ as

$$f_n^m(E_\theta) = \frac{\sqrt{n+m+1/2}}{2^m \Gamma(n+m+1)} \sqrt{\Gamma(n+1)\Gamma(n+2m+1)} \; H_n^m(\xi;\gamma), \tag{3.9}$$

transforms the recursion relation (3.8) into the following

$$\left(\gamma+\tfrac{1}{2}\right)^2 H_n^m = \left(n+m+\tfrac{1}{2}\right)^2 H_n^m - \xi \tfrac{n+m}{(n+m+1/2)} H_{n-1}^m - \xi \tfrac{(n+1)(n+2m+1)}{(n+m+1/2)(n+m+1)} H_{n+1}^m \tag{3.10}$$

The solutions of this recursion relation (for fixed $\xi$ and $m$) are functions in $\gamma$ that are defined modulo an arbitrary non-singular factor that depends on $\gamma$ but otherwise independent of $n$. If we choose the standard normalization by taking the initial seed as $H_0^m = 1$ then this gives $H_n^m(\xi;\gamma)$ as a polynomial of degree $n$ in $\left(\gamma+\tfrac{1}{2}\right)^2$. These polynomials, as far as we know, were never studied in sufficient details before. Since they are associated with the electric dipole potential $\xi \frac{\cos\theta}{r^2}$ we refer to them as the "*dipole polynomials*" of the first kind. The "*dipole polynomials*" of the second kind will be defined below in Sec. V when we study the case of a neural molecule (i.e., when the electric dipole has zero net charge). Now, for a given $m$ and $\xi$, the recursion relation (3.10) together with the initial value $H_0^m = 1$ and the definition $H_{-1}^m \equiv 0$ determine completely the set of all polynomials $\{H_n^m(\xi;\gamma)\}$. A generalized version of these polynomials, where the integer $m$ is replaced by a continuous parameter $\mu$, is introduced and studied in Appendix B. The angular component of the wavefunction, $\Omega(\theta,\phi) = \Theta(\theta)\Phi(\phi)$, could now be written as the $L^2$–series sum

$$\Omega(\theta,\phi) = \sum_{m=M}^{\infty} A_m^{\xi,\gamma} \sum_{n=0}^{\infty} \frac{(n+m+1/2)\Gamma(n+2m+1)}{2^{2m}\Gamma(n+m+1)^2 / \Gamma(n+1)} H_n^m(\xi;\gamma)(1-x^2)^{\frac{m}{2}} P_n^{(m,m)}(x) e^{\pm i m\phi} \tag{3.11}$$

where $M$ is a non-negative integer that depends on the value of the dipole moment $\xi$. It is the smallest integer such that $\hat{\xi}_M \geq \xi$. $A_m^{\xi,\gamma}$ is a normalization constant that depends on the physical parameters but, otherwise, independent of $n$. To make $\Omega(\theta,\phi)$ $\gamma$-normalizable, we write $A_m^{\xi,\gamma} = \sqrt{\frac{1}{2\pi} \rho^m(\xi;\gamma)}$, where $\rho^m(\xi;\gamma)$ is the weight (density) function associated with these dipole polynomials $\{H_n^m\}$. That is,

$$\int \rho^m(\xi;\gamma) H_n^m(\xi;\gamma) H_{n'}^m(\xi;\gamma) d\gamma = \frac{2^{2m}\Gamma(n+m+1)^2/\Gamma(n+1)}{(n+m+1/2)\Gamma(n+2m+1)} \delta_{nn'}. \tag{3.12}$$

It should be obvious from Eq. (3.6) that the diagonal representation, where $H_\theta |\chi_n\rangle = E_\theta |\chi_n\rangle$, is obtained if and only if $\xi = 0$ and

$$\gamma = \begin{cases} \ell & ; \gamma \geq 0 \\ -\ell - 1 & ; \gamma \leq -1 \end{cases} \tag{3.13}$$

where $\ell = n + m = 0, 1, 2, \ldots$ This means that a diagonal representation is obtained only in the absence of the electric dipole. This might be the reason behind the curious absence of an *exact analytic* solution to this problem in the literature. This also demonstrates the advantage and utility of the tridiagonal representation, which is a signature of our "Tridiagonal Physics" program [7]. The recursion (3.10) shows that the discrete values of $\gamma$ given by Eq. (3.13) are obtained not only in the absence of the electric dipole ($\xi = 0$) but also asymptotically for large $n$ or $m$. It is also important to note that the diagonal representation we are referring to here is associated with the angular operator $H_\theta$. That is, $(H_\theta)_{nn'} = E_\theta \delta_{nn'}$. One should not confuse this with the discrete bound states spectrum, which is associated with the diagonal representation of the total Hamiltonian $H$ (i.e., $H_{nn'} = E\delta_{nn'}$). Consequently, it is neither required nor necessary to impose the constraints (3.13) or $\xi = 0$ on the bound states solution. That is, for bound states it is neither required that $\xi$ vanishes nor that $\gamma$ (equivalently, $E_\theta$) be quantized as shown in Eq. (3.13). These points will be re-emphasized when we construct the radial component of the wavefunction and obtain the bound states energy spectrum in the following section.

## IV. RADIAL COMPONENT OF THE WAVEFUNCTION

The radial wavefunction $R(r)$ could be taken as an element in the space spanned by the following $L^2$ functions that are compatible with the domain of the radial component of the Hamiltonian $H_r$:

$$\phi_k(y) = b_k y^\tau e^{-y/2} L_k^\sigma(y), \tag{4.1}$$

where $y = \lambda r$, $k = 0,1,2,...$, and $L_k^\sigma(y)$ is the Laguerre polynomial of degree $k$. The real parameter $\lambda$ is positive with an inverse length dimension (i.e., it is a length scale parameter). On the other hand, the dimensionless parameters $\tau > 0$ and $\sigma > -1$. The normalization constant, $b_k = \sqrt{\lambda \Gamma(k+1)/\Gamma(k+\sigma+1)}$, is chosen to conform with the orthogonality property of the Laguerre polynomials. Using the differential equation (A.8) and differential formula (A.9) for the Laguerre polynomials in Appendix A, we obtain

$$\begin{aligned}\frac{d^2\phi_k}{dr^2} &= \lambda^2 \left[ -\frac{k}{y}\left(1 + \frac{\sigma+1-2\tau}{y}\right) + \frac{\tau(\tau-1)}{y^2} - \frac{\tau}{y} + \frac{1}{4} \right]\phi_k \\ &\quad - \lambda^2 \frac{(k+\sigma)(2\tau-\sigma-1)}{y^2} \frac{b_k}{b_{k-1}} \phi_{k-1}\end{aligned} \tag{4.2}$$

Therefore, the action of the radial differential wave operator in Eq. (2.4c) on the basis element (4.1) gives the following

$$\begin{aligned}(H_r - E)|\phi_k\rangle &= \frac{\lambda^2}{2}\left[ \frac{k}{y}\left(1 + \frac{\sigma+1-2\tau}{y}\right) + \frac{\left(\gamma+\tfrac{1}{2}\right)^2 - \left(\tau-\tfrac{1}{2}\right)^2}{y^2} + \frac{\tau}{y} - \frac{1}{4} \right. \\ &\quad \left. + \frac{2}{\lambda^2}(V_r - E) \right]|\phi_k\rangle + \frac{\lambda^2}{2}\frac{(k+\sigma)(2\tau-\sigma-1)}{y^2}\frac{b_k}{b_{k-1}}|\phi_{k-1}\rangle\end{aligned} \tag{4.3}$$

The recurrence relation (A.6) and orthogonality formula (A.10) for the Laguerre polynomials show that a tridiagonal matrix representation $\langle \phi_k | H_r - E | \phi_{k'} \rangle$ is possible with $V_r \sim y^{-1}$ if and only if $\sigma = 2\tau - 1$ and $\left(\tau - \tfrac{1}{2}\right)^2 = \left(\gamma + \tfrac{1}{2}\right)^2$. That is,

$$\tau = \begin{cases} \gamma + 1 & ; \gamma \geq 0 \\ -\gamma & ; \gamma \leq -1 \end{cases} \tag{4.4}$$

which makes $\tau$ always greater than or equal to +1 and $\sigma = \pm(2\gamma+1)$ for $\pm\gamma > 0$.

We expand the radial component of the wavefunction in the basis (4.1) as $R(r) = \sum_{k=0}^\infty g_k^\gamma(E)\phi_k(y)$ and substitute this in the wave equation (2.4c). Employing the action of the wave operator on the basis as given by Eq. (4.3) and using the recurrence relation and orthogonality formula of the Laguerre polynomials, we obtain the following tridiagonal matrix representation of the radial wave operator

$$\frac{2}{\lambda^2}\langle\phi_k|H_r-E|\phi_{k'}\rangle = \left[(2k+\sigma+1)\left(\frac{1}{4}-\frac{2E}{\lambda^2}\right)+2\frac{Z}{\lambda}\right]\delta_{k,k'}$$
$$+\left(\frac{1}{4}+\frac{2E}{\lambda^2}\right)\left[\sqrt{k(k+\sigma)}\,\delta_{k,k'+1}+\sqrt{(k+1)(k+\sigma+1)}\,\delta_{k,k'-1}\right] \qquad (4.5)$$

Therefore, the resulting three-term recursion relation for the expansion coefficients of the radial wavefunction becomes

$$\left[(2k+\sigma+1)\frac{\zeta_-}{\zeta_+}-\frac{2Z/\lambda}{\zeta_+}\right]g_k^\gamma - \sqrt{k(k+\sigma)}\,g_{k-1}^\gamma - \sqrt{(k+1)(k+\sigma+1)}\,g_{k+1}^\gamma = 0, \qquad (4.6)$$

where $\zeta_\pm = \frac{2E}{\lambda^2}\pm\frac{1}{4}$. Rewriting this recursion in terms of the polynomials $S_k^\gamma(Z;E) = \sqrt{\Gamma(k+\sigma+1)/\Gamma(k+1)}\,g_k^\gamma(E)$, we obtain a more familiar recursion relation as follows

$$2\left[\left(k+\frac{\sigma+1}{2}+2\frac{Z}{\lambda}\right)\cos\varphi - 2\frac{Z}{\lambda}\right]S_k^\gamma = (k+\sigma)S_{k-1}^\gamma + (k+1)S_{k+1}^\gamma. \qquad (4.7)$$

where $\cos\varphi = \zeta_-/\zeta_+ = \frac{8E-\lambda^2}{8E+\lambda^2}$. We compare this three-term recursion relation to that of a special case of the Pollaczek polynomials, $P_k^\mu(\nu;z)$ [8]:

$$2\left[(k+\mu+\nu)z-\nu\right]P_k^\mu(\nu;z) = (k+2\mu-1)P_{k-1}^\mu(\nu;z) + (k+1)P_{k+1}^\mu(\nu;z), \qquad (4.8)$$

where $\mu > 0$, $\nu$ is real, $z = \cos\varphi$ and $\pi > \varphi > 0$. These polynomials could be written in terms of the hypergeometric function as follows

$$P_k^\mu(\nu;z) = \frac{\Gamma(k+2\mu)}{\Gamma(k+1)\Gamma(2\mu)}e^{ik\varphi}\,_2F_1\left(-k,\mu+i\nu\eta;2\mu;1-e^{-2i\varphi}\right), \qquad (4.9)$$

where $\eta = \frac{\cos\varphi-1}{\sin\varphi} = -\tan\frac{\varphi}{2}$. The associated orthogonality relation is

$$\int_{-1}^{+1}\rho^\mu(\nu;z)P_k^\mu(\nu;z)P_{k'}^\mu(\nu;z)dz = \frac{\Gamma(k+2\mu)}{(k+\mu+\nu)\Gamma(k+1)}\delta_{kk'}, \qquad (4.10)$$

where the weight function $\rho^\mu(\nu;z) = \frac{1}{\pi}(2\sin\varphi)^{2\mu-1}e^{(2\varphi-\pi)\nu\eta}|\Gamma(\mu+i\nu\eta)|^2$. Therefore, we can write $S_k^\gamma(Z;E) \propto P_k^{\frac{\sigma+1}{2}}\left(2\frac{Z}{\lambda};\cos\varphi\right)$. However, this is valid only within the permissible range of values of the parameters. This means that the solution obtained as such is valid only for $E > 0$ (i.e., for the continuum scattering states). Thus for the *continuum* case, we obtain the following $L^2$–series solutions for the radial component of the wavefunction

$$R(r) = B_\gamma^E(\lambda r)^{\frac{\sigma+1}{2}}e^{-\lambda r/2}\sum_{k=0}^\infty \frac{\Gamma(k+1)}{\Gamma(k+\sigma+1)}P_k^{\frac{\sigma+1}{2}}\left(2\frac{Z}{\lambda},\cos\varphi\right)L_k^\sigma(\lambda r), \qquad (4.11)$$

where $\varphi$ is energy dependent as given above by $\varphi(E) = \cos^{-1}\left(\frac{8E-\lambda^2}{8E+\lambda^2}\right)$ and $\sigma = \pm(2\gamma+1)$ for $\pm\gamma > 0$. The normalization constant $B_\gamma^E$ depends on $\gamma$ and the energy but, otherwise, independent $k$. To make $R(r)$ energy-normalized we write $B_\gamma^E = \sqrt{\rho^{\frac{\sigma+1}{2}}\left(2\frac{Z}{\lambda};\cos\varphi\right)/E}$.

To obtain the discrete energy representation, we impose the diagonalization constraint on the tridiagonal matrix representation (4.5) giving $\lambda^2 = -8E$ (i.e., the energy must be negative for real representation) and the following

$$\lambda_k = -2Z/\left(k+\frac{\sigma+1}{2}\right) = \begin{cases} -2Z/(k+\gamma+1) & ;\gamma \geq 0 \\ -2Z/(k-\gamma) & ;\gamma \leq -1 \end{cases} \qquad (4.12)$$

which requires that $qZ < 0$ since $\lambda$ must be positive. That is, bound states exist only for an attractive Coulomb potential (i.e., positively ionized molecule). Thus, the discrete bound states energy spectrum is

$$E_k = -Z^2 \big/ 2\left(k + \tfrac{\sigma+1}{2}\right)^2 = \begin{cases} -Z^2/2(k+\gamma+1)^2 & ;\gamma \geq 0 \\ -Z^2/2(k-\gamma)^2 & ;\gamma \leq -1 \end{cases} \tag{4.13}$$

The corresponding radial component of the discrete bound states wavefunction is

$$R_k(r) = \phi_k(y) = \sqrt{\tfrac{\lambda_k \Gamma(k+1)}{\Gamma(k+\sigma+1)}} (\lambda_k r)^{\tfrac{\sigma+1}{2}} e^{-\lambda_k r/2} L_k^\sigma(\lambda_k r), \tag{4.14}$$

In the following section, we study in details the special and singular case where $Z = 0$ (i.e., the Coulomb-free interaction). It corresponds to the pure dipole interaction where the molecule is neutral.

## V. PURE DIPOLE INTERACTION

Now, if the molecule is not ionized (equivalently, $Z = 0$) but still possesses an electric dipole moment, then the interaction will be Coulomb-free and $V_r = 0$. In this case, the angular component of the wave function for the continuum states does not change from that given by Eq. (3.11). However, the three-term recursion relation (4.7) associated with the radial component of the wavefunction becomes

$$2\left(k + \tfrac{\sigma+1}{2}\right) \cos\varphi \, S_k^\gamma = (k+\sigma) S_{k-1}^\gamma + (k+1) S_{k+1}^\gamma. \tag{5.1}$$

This is the recursion relation of the ultra-spherical (Gegenbauer) polynomials $C_k^{\tfrac{\sigma+1}{2}}(\cos\varphi)$ where the corresponding weight function is $(\sin\varphi)^\sigma$ [8]. Therefore, the *continuum* radial wavefunction becomes

$$R(r) = B_\gamma^E (\lambda r)^{\tfrac{\sigma+1}{2}} e^{-\lambda r/2} \sum_{k=0}^\infty \frac{\Gamma(k+1)}{\Gamma(k+\sigma+1)} C_k^{\tfrac{\sigma+1}{2}}(\cos\varphi) L_k^\sigma(\lambda r), \tag{5.2}$$

where, again, $\sigma = \pm(2\gamma+1)$ for $\pm\gamma > 0$.

Next, we study the solution space for *bound* states, where the electron is captured by the molecule. For pure electric dipole interaction (i.e., when $V_r = 0$), Eq. (2.4c) turns into the radial Schrödinger equation for the inverse square potential. It has been well established that bound states in this $r^{-2}$ singular potential could be supported only if the value of its dimensionless coupling parameter is less than or equal to the critical value $-\tfrac{1}{4}$ [9]. Therefore, we write $2E_\theta = -\tfrac{1}{4} - \omega^2$ where $\omega$ is real. This is equivalent to the definition given for $E_\theta$ above in Sec. III but with $\gamma \to i\omega - \tfrac{1}{2}$. If we now expand the angular component of the wavefunction in the $L^2$ basis (3.2), then the new expansion coefficients will be written in terms of orthogonal polynomials, $G_n^m(\xi;\omega)$, that satisfy a different three-term recursion relation that reads as follows

$$\omega^2 G_n^m = -\left(n+m+\tfrac{1}{2}\right)^2 G_n^m + \xi \tfrac{n+m}{(n+m+1/2)} G_{n-1}^m + \xi \tfrac{(n+1)(n+2m+1)}{(n+m+1/2)(n+m+1)} G_{n+1}^m \tag{5.3}$$

We refer to these as the "*dipole polynomials of the second kind*". By comparing (5.3) with the recursion (3.10) we can relate these two kinds by $G_n^m(\xi;\omega) = H_n^m\left(\xi; i\omega - \tfrac{1}{2}\right)$. This recursion relation shows that if $\xi$ becomes too small then reality of the representation will be compromised. Following the same argument that was made above following Eq. (3.8), we require that the dipole moment should be greater than the critical value $\hat{\xi}_m$ for some integer $m$. In other words, for a given physical dipole moment $\xi$, the azimuthal quantum number should not exceed the largest integer $M$ such that $\hat{\xi}_M \leq \xi$. Therefore, the angular component of the wavefunction becomes

$$\Omega(\theta,\phi) = \sum_{m=0}^{M} A_m^{\xi,\gamma} \sum_{n=0}^{\infty} \frac{(n+m+1/2)\Gamma(n+2m+1)}{2^{2m}\Gamma(n+m+1)^2/\Gamma(n+1)} H_n^m(\xi;\gamma)(1-x^2)^{\frac{m}{2}} P_n^{(m,m)}(x) e^{\pm i m\phi} \quad (5.4)$$

One should also note that the constraint on the quantum number $m$ to be within the permissible range $0 \leq m \leq M$ is analogous to the restriction on the azimuthal quantum number (in spherically symmetric problems) to be in the range $-\ell, -\ell+1, ..., \ell-1, \ell$. Moreover, this constraint makes at least one of the zeros of the polynomial $G_n^m(\xi;\omega)$ on the $\omega$-axis real.

It has also been well established long ago that the electric dipole moment $\xi$ must exceed a critical value so that the solution space for the electron-molecule bound states becomes non-empty [2]. Mathematically, it could be obvious that this constraint is closely related to the critical value of the coupling parameter of the inverse square potential mentioned above. To obtain this critical dipole value, we investigate Eq. (3.8) with $\gamma \to i\omega - \frac{1}{2}$, which could now be written as the eigenvalue equation $h|f\rangle = \omega^2 |f\rangle$, where $h$ is the tridiagonal symmetric matrix

$$h = \begin{pmatrix} A_0 & B_0 & & & & \\ B_0 & A_1 & B_1 & & \mathbf{0} & \\ & B_1 & A_2 & B_2 & & \\ & & \times & \times & \times & \\ & \mathbf{0} & & \times & \times & \times \\ & & & & \times & \times \end{pmatrix} \quad (5.5)$$

with $A_n = -\left(n+m+\frac{1}{2}\right)^2$ and $B_n = \xi\sqrt{\frac{(n+1)(n+2m+1)}{(n+m+1)^2 - 1/4}}$. Therefore, the determinant of the matrix $h - \omega^2 I$ must vanish, where $I$ is the identity matrix. This requirement is necessary so that the kernel of this matrix operator become non-singular (i.e., to prevent the solution space $\{f_n^m\}$ from becoming empty). Consequently, this translates into a condition on the electric dipole parameter $\xi$ that depends on $m$ and $\omega$. The critical value of $\xi$ is the smallest value that satisfies this condition. It corresponds to the case when the eigenvalue of the matrix $h$ vanishes (i.e. when $\omega = 0$) and we refer to it as $\hat{\xi}_m$. For the lowest few azimuthal quantum numbers $m$, Table 2 shows a sequence of these critical values of $\xi$ for an $N$-dimensional matrix $h$ with $N = 2, 3, .., 12$. It is evident that the sequence converges rapidly with $N$ for the given choice of significant digits. In fact, for each $N$ one finds a set of $2K$ zeros, $\{\pm \xi_i\}_{i=1}^{K}$, of the determinant where $K = \frac{N}{2}$ or $K = \frac{N-1}{2}$ if $N$ is even or odd, respectively. The smallest positive zero (for large enough $N$) is the critical value of the dipole moment $\hat{\xi}_m$. Table 3 lists these zeros for several values of the azimuthal number $m$ and for $N = 10$. These zeros converge to the values given in Table 4 for large *N*. The values obtained in this Table agree with those already reported long ago by Crawford in [10].

Now, we turn attention to the radial component of wavefunction for bound states. The action of the radial wave operator (2.4c) with $V_r = 0$ and $2E_\theta = -\frac{1}{4} - \omega^2$ on the basis element (4.1) gives the following

$$(H_r - E)|\phi_k\rangle = \frac{\lambda^2}{2}\left[\frac{k}{y}\left(1 + \frac{\sigma+1-2\tau}{y}\right) - \frac{\omega^2 + (\tau-\frac{1}{2})^2}{y^2} + \frac{\tau}{y} - \frac{1}{4} - \frac{2E}{\lambda^2}\right]|\phi_k\rangle$$
$$+ \frac{\lambda^2}{2}\frac{(k+\sigma)(2\tau-\sigma-1)}{y^2}\frac{b_k}{b_{k-1}}|\phi_{k-1}\rangle \quad (5.6)$$

The recurrence relation (A.6) and orthogonality formula (A.10) for the Laguerre polynomials show that a tridiagonal matrix representation $\langle\phi_k|H_r - E|\phi_{k'}\rangle$ is possible if and only if $\sigma = 2\tau - 2$ and $\lambda^2 = -8E$. Thus, the basis becomes energy dependent via the parameter $\lambda$ and we obtain the following tridiagonal matrix elements for the radial wave operator in this basis

$$\langle\phi_k|H_r - E|\phi_{k'}\rangle = 4E\left[\omega^2 + \left(\frac{\sigma+1}{2}\right)^2 + k - (2k+\sigma+1)\left(k+\frac{\sigma}{2}+1\right)\right]\delta_{k,k'}$$
$$+ 4E\left[\left(k+\frac{\sigma}{2}\right)\sqrt{k(k+\sigma)}\delta_{k,k'+1} + \left(k+\frac{\sigma}{2}+1\right)\sqrt{(k+1)(k+\sigma+1)}\delta_{k,k'-1}\right] \quad (5.7)$$

Thus, the radial wave equation becomes equivalent to a three-term recursion relation for the expansion coefficients, $g_k^\omega(E)$, of the wavefunction. Again, rewriting this recursion in terms of the polynomials $S_k(\omega, E) = \sqrt{\frac{\Gamma(k+\sigma+1)}{\Gamma(k+1)}}g_k^\omega(E)$, we obtain the following

$$\omega^2 S_k = \left[(k+\sigma+1)\left(k+\frac{\sigma}{2}+1\right) + k\left(k+\frac{\sigma}{2}\right) - \left(\frac{\sigma+1}{2}\right)^2\right]S_k$$
$$- k\left(k+\frac{\sigma}{2}\right)S_{k-1} - (k+\sigma+1)\left(k+\frac{\sigma}{2}+1\right)S_{k+1} \quad (5.8)$$

One should observe the curious absence of the energy from this recursion relation. This implies that the dependence of $S_k(\omega, E)$ on energy is via an overall factor, which is independent of *k*, say $F_\omega^\sigma(E)$. The source of this factorization comes from (5.7), where all elements of the matrix representation of the radial wave operator have the common factor 4*E*. This means that the wave equation, in this representation, is satisfied independently of any value of the energy as long as it is negative (due to $\lambda^2 = -8E$). Consequently, this property has a dramatic implication on the bound states energy spectrum. It implies that for any choice of negative energy a bound state could be supported. That is, the energy spectrum is continuous and, in fact, it is the complete semi-infinite negative real line. However, the diagonal constraint on the representation (5.7) dictates that $E = 0$. These observations, concerning the peculiar behavior of the bound states energy spectrum, have already been reported and analyzed in the literature for the inverse square potential. See, for example, [9] and references therein. Regularization procedures [9,11] and self-adjoint extensions of the Hamiltonian [12] were introduced to handle these irregularities (anomalies) in the spectrum.

Now, Eq. (5.8) is a special case of the three-term recursion relation of the continuous dual Hahn orthogonal polynomials, $Q_k^\mu(x; a, b)$, where *x* is real and $\mu$, *a*, *b* are positive except for a possible pair of complex conjugates with positive real parts [13]. The general recursion relation for these polynomials reads as follows

$$x^2 Q_k^\mu = \left[(k+\mu+a)(k+\mu+b) + k(k+a+b-1) - \mu^2\right] Q_k^\mu \\ -k(k+a+b-1)Q_{k-1}^\mu - (k+\mu+a)(k+\mu+b)Q_{k+1}^\mu \quad (5.9)$$

and with the standard initial seed, $Q_0^\mu = 0$. These polynomials could be written in terms of the generalized hypergeometric function as follows

$$Q_k^\mu(x;a,b) = {}_3F_2\left(\begin{matrix}-k,\mu+ix,\mu-ix\\ \mu+a,\mu+b\end{matrix}\Big|1\right). \quad (5.10)$$

The associated orthogonality relation is

$$\int_0^\infty \rho^\mu(x) Q_n^\mu(x;a,b) Q_m^\mu(x;a,b)\, dx = \frac{\Gamma(n+1)\Gamma(n+a+b)}{\Gamma(n+\mu+a)\Gamma(n+\mu+b)}\delta_{nm}, \quad (5.11)$$

where the weight function is $\rho^\mu(x) = \frac{1}{2\pi}\left|\frac{\Gamma(\mu+ix)\Gamma(a+ix)\Gamma(b+ix)}{\Gamma(\mu+a)\Gamma(\mu+b)\Gamma(2ix)}\right|^2$. Comparing (5.9) with (5.8), we can write

$$g_k^\omega(E) \propto \sqrt{\frac{\Gamma(k+1)}{\Gamma(k+\sigma+1)}}\, Q_k^{\frac{\sigma+1}{2}}\left(\omega; \frac{\sigma+1}{2}, \frac{1}{2}\right). \quad (5.12)$$

Thus, the expansion coefficients $\{g_k^\omega(E)\}_{k=0}^\infty$ are defined modulo the *arbitrary* factor $F_\omega^\sigma(E)$ that depends on $\omega$ and $E$. Finally, the radial component of the wavefunction for the bound state at energy $E$ becomes

$$R(E,r) = F_\omega^\sigma (2\eta r)^{\frac{\sigma+1}{2}} e^{-\eta r} \sum_{k=0}^\infty \frac{\Gamma(k+1)}{\Gamma(k+\sigma+1)} Q_k^{\frac{\sigma+1}{2}}\left(\omega; \frac{\sigma+1}{2}, \frac{1}{2}\right) L_k^\sigma(2\eta r), \quad (5.13)$$

where the energy dependent wave number $\eta$ is defined by $E = -\frac{1}{2}\eta^2$.

## VI. SUMMARY AND CONCLUSION

In the conventional diagonal representation of the eigenvalue wave equation, $H|\phi_n\rangle = E|\phi_n\rangle$, the noncentral electric dipole potential $\cos\theta/r^2$ does not belong to any of the known exactly solvable potentials. However, using the "Tridiagonal Physics" formulation [7], we were able to construct an exact solution space for the single particle Hamiltonian associated with this noncentral potential. Consequently, we obtained an exact analytic solution for this problem that models the interaction of an excess electron with a frozen dipolar molecule. The problem is reduced to finding solutions of the resulting three-term recursion relation for the expansion coefficient of the wavefunction in a suitable square integrable basis. The solutions of such three-term recursion relations are usually expressed in terms of variants of the classical orthogonal polynomials. However, in this case the three-term recursion relation associated with the angular component of the wavefunction generated polynomials that were not studied rigorously in the past. Adopting the standard normalization, these orthogonal polynomials were completely defined. Since they are associated with the electric dipole potential, we referred to them as the "*dipole polynomials*". In a proper mathematical setting, we hope to be able in the near future to obtain the density (weight) function associated with these polynomials as given by Eq. (3.12). On the other hand, the radial component of the wavefunction was also obtained analytically for the scattering as well as the bound states. These were written in terms of the Gegenbauer polynomials and the continuous dual Han polynomials, respectively. Therefore, the noncentral electric dipole potential $\cos\theta/r^2$ becomes now a new element in the class of exactly solvable potentials. The calculation we have performed supports the existence of a dipole-bound anion for dipole moments higher than a certain critical value. The critical value occurs when the ground state energy reaches the zero value. These critical dipole moments were evaluated not only for the ground state, where they agree with already known and experimentally verified results, but also for excited states as well. Unfortunately, there are no experimental data available for the excited states (that we are aware of) and no comparison could be made with our numerical results.

Additionally, we also obtained analytic solution for the case of ionized dipolar molecule. That is, we presented $L^2$ series solutions for the discrete and continuum states associated with the noncentral potential $V(r,\theta) = Z/r + \xi\cos\theta/r^2$, where $\xi$ is the electric dipole moment and $Z$ is its net charge. The angular component of the wavefunction for this case is written in terms of the orthogonal *dipole polynomials of the first kind*, whereas, the radial component (for the continuum case) is written in terms of the Pollaczek polynomials. The discrete bound states energy spectrum is given by Eq. (4.13).

## ACKNOWLEDGMENTS

I am grateful to H. Bahlouli for stimulating discussions and critical review of the manuscript. I am also indebted to M. E. H. Ismail for fruitful correspondence. The support provided by KFUPM library and the American University of Sharja in literature survey are highly appreciated.

# APPENDIX A: THE LAGUERRE AND JACOBI POLYNOMIALS

The following are useful identities, formulas, and relations associated with the Laguerre and Jacobi polynomials. They are found in most textbooks and monographs on orthogonal polynomials [8] but listed here for ease of reference.

(1) The Jacobi polynomials $P_n^{(\mu,\nu)}(x)$, where $x \in [-1,+1]$ and $\mu > -1, \nu > -1$:

$$\left(\frac{1 \pm x}{2}\right) P_n^{(\mu,\nu)} = \frac{2n(n+\mu+\nu+1) + (\mu+\nu)(\frac{\mu+\nu}{2} \pm \frac{\nu-\mu}{2}+1)}{(2n+\mu+\nu)(2n+\mu+\nu+2)} P_n^{(\mu,\nu)}$$
$$\pm \frac{(n+\mu)(n+\nu)}{(2n+\mu+\nu)(2n+\mu+\nu+1)} P_{n-1}^{(\mu,\nu)} \pm \frac{(n+1)(n+\mu+\nu+1)}{(2n+\mu+\nu+1)(2n+\mu+\nu+2)} P_{n+1}^{(\mu,\nu)} \quad (A.1)$$

$$P_n^{(\mu,\nu)}(x) = \frac{\Gamma(n+\mu+1)}{\Gamma(n+1)\Gamma(\mu+1)} {}_2F_1(-n, n+\mu+\nu+1; \mu+1; \tfrac{1-x}{2}) = (-)^n P_n^{(\nu,\mu)}(-x) \quad (A.2)$$

$$\left\{(1-x^2)\frac{d^2}{dx^2} - [(\mu+\nu+2)x + \mu-\nu]\frac{d}{dx} + n(n+\mu+\nu+1)\right\} P_n^{(\mu,\nu)}(x) = 0 \quad (A.3)$$

$$(1-x^2)\frac{d}{dx} P_n^{(\mu,\nu)} = -n\left(x + \frac{\nu-\mu}{2n+\mu+\nu}\right) P_n^{(\mu,\nu)} + 2\frac{(n+\mu)(n+\nu)}{2n+\mu+\nu} P_{n-1}^{(\mu,\nu)} \quad (A.4)$$

$$\int_{-1}^{+1} (1-x)^\mu (1+x)^\nu P_n^{(\mu,\nu)}(x) P_m^{(\mu,\nu)}(x) dx = \frac{2^{\mu+\nu+1}}{2n+\mu+\nu+1} \frac{\Gamma(n+\mu+1)\Gamma(n+\nu+1)}{\Gamma(n+1)\Gamma(n+\mu+\nu+1)} \delta_{nm} \quad (A.5)$$

(2) The Laguerre polynomials $L_n^\nu(x)$, where $x \in [0, \infty]$ and $\nu > -1$:

$$xL_n^\nu = (2n+\nu+1)L_n^\nu - (n+\nu)L_{n-1}^\nu - (n+1)L_{n+1}^\nu \quad (A.6)$$

$$L_n^\nu(x) = \frac{\Gamma(n+\nu+1)}{\Gamma(n+1)\Gamma(\nu+1)} {}_1F_1(-n; \nu+1; x) \quad (A.7)$$

$$\left[x\frac{d^2}{dx^2} + (\nu+1-x)\frac{d}{dx} + n\right] L_n^\nu(x) = 0 \quad (A.8)$$

$$x\frac{d}{dx} L_n^\nu = nL_n^\nu - (n+\nu)L_{n-1}^\nu \quad (A.9)$$

$$\int_0^\infty x^\nu e^{-x} L_n^\nu(x) L_m^\nu(x) dx = \frac{\Gamma(n+\nu+1)}{\Gamma(n+1)} \delta_{nm} \quad (A.10)$$

# APPENDIX B: THE GENERALIZED DIPOLE POLYNOMIAL

The orthogonal dipole polynomials (of the first kind) defined by the three-term recursion relation (3.10) and Eq. (3.12) could be generalized for non-integral indices. If we replace the non-negative integer *m* by the continuous parameter $\mu - \tfrac{1}{2}$, then relation (3.10) becomes

$$xH_n^\mu(\xi; x) = (n+\mu)^2 H_n^\mu(\xi; x) - \xi\frac{n+\mu+1/2}{n+\mu} H_{n-1}^\mu(\xi; x) - \xi\frac{(n+1)(n+2\mu)}{(n+\mu)(n+\mu+1/2)} H_{n+1}^\mu(\xi; x), \quad (B.1)$$

where *x* and $\mu$ are real and positive. Starting with the standard seed $H_0^\mu = 1$, this relation specifies these polynomials recursively for all degrees in *x*:

$$H_0^\mu(\xi; x) = 1, \quad (B.2a)$$

$$H_1^\mu(\xi; x) = -\frac{\mu+1/2}{2\xi}(x - \mu^2), \quad (B.2b)$$

$$H_2^\mu(\xi;x) = \frac{(\mu+1)(\mu+3/2)}{8\xi^2}\left\{(x-\mu^2)\left[x-(\mu+1)^2\right] - 2\xi^2 \frac{\mu+3/2}{(\mu+1/2)(\mu+1)}\right\}, \tag{B.2c}$$

..........
..........

$$H_n^\mu(\xi;x) = \frac{(n+\mu-1)(n+\mu-1/2)}{\xi n(n+2\mu-1)}\left\{\left[(n-1+\mu)^2 - x\right]H_{n-1}^\mu(\xi;x) - \xi\frac{n+\mu-1/2}{n+\mu-1}H_{n-2}^\mu(\xi;x)\right\}. \tag{B.2d}$$

For these generalized polynomials, the critical dipole values become a function of the continuous parameter as $\hat{\xi}(\mu)$. A simple method to obtain an accurate evaluation of this characteristic function goes as follows. We rewrite the recursion (B.1) for $x = 0$ in terms of the polynomials $f_n^\mu$, which are related to $H_n^\mu$ by Eq. (3.9) with $m \to \mu - \frac{1}{2}$, as follows

$$(n+\mu)^2 f_n^\mu = \xi\sqrt{\frac{n(n+2\mu-1)}{(n+\mu-1)(n+\mu)}}f_{n-1}^\mu + \xi\sqrt{\frac{(n+1)(n+2\mu)}{(n+\mu)(n+\mu+1)}}f_{n+1}^\mu. \tag{B.3}$$

Defining a new polynomial $q_n^\mu = \frac{1}{\mu}(n+\mu)f_n^\mu$, this relation becomes

$$\kappa q_n^\mu(\kappa) = \alpha_{n-1}q_{n-1}^\mu(\kappa) + \alpha_n q_{n+1}^\mu(\kappa), \tag{B.4}$$

where $\kappa = \xi^{-1}$ and $\alpha_n = \frac{1}{(n+\mu)(n+\mu+1)}\sqrt{\frac{(n+1)(n+2\mu)}{(n+\mu)(n+\mu+1)}}$. Equation (B.4) is an eigenvalue equation for the tridiagonal symmetric matrix $T_{nm} = \alpha_n\delta_{n,m-1} + \alpha_m\delta_{n,m+1}$ with eigenvalue $\kappa$ and corresponding eigenvector $\{q_n\}$. For a given $\mu$, it is straightforward and simple to calculate the eigenvalues $\{\kappa_n(\mu)\}_{n=0}^{N-1}$ of an $N \times N$ finite version of this tridiagonal matrix using almost any standard computational package. The smallest positive value in the corresponding set $\{\hat{\xi}_n(\mu)\}_{n=0}^{N-1}$ is the critical dipole moment for that given $\mu$. The rest are higher order critical dipole moments that are associated with the excited states indexed by $n$ for that value of $\mu$. Figure 1 is a plot of the critical dipole moment functions $\{\hat{\xi}_n(\mu)\}_{n=0}^{10}$ associated with the tridiagonal matrix $T$ whose dimension $N = 100$ and for $\mu \in [0,30]$. All Tables in this work could be reproduced, to the desired accuracy, as special cases of these generalized dipole polynomials with $\mu = m$. In a future work under a proper mathematical setting, we hope to be able to obtain the density (weight) function $\rho^\mu(\xi;x)$ associated with these generalized dipole polynomials as follows

$$\int_0^\infty \rho^\mu(\xi;x)H_n^\mu(\xi;x)H_m^\mu(\xi;x)dx = \frac{2^{2\mu-1}\Gamma(n+\mu+1/2)^2}{(n+\mu)\Gamma(n+2\mu)\Gamma(n+1)}\delta_{nm}. \tag{B.5}$$

## TABLE CAPTIONS

**Table 1**: Critical values (in atomic units) of the electric dipole moment $\{\hat{\xi}_m\}_{m=0}^{10}$.

**Table 2**: For the lowest few azimuthal quantum numbers $m$, we show a sequence of values of the dipole moment (in atomic units) converging to the critical value $\hat{\xi}_m$ as the dimension $N$ of the matrix representation of the angular wave operator increases.

**Table 3**: The zeros (in atomic units) of the determinant of the matrix representation of the angular component of the wave operator (5.5) for $N = 10$. The smallest positive zero is the critical value of the dipole moment for the given state indexed by $m$.

**Table 4**: The limits to which the values of the dipole moments given in Table 3 converge for large values of $N$. We took $N = 50$.

**Table 1**

| $m$ | $\hat{\xi}_m$ (a.u.) |
|---|---|
| 0 | 0.639314877199981 |
| 1 | 3.791967926767455 |
| 2 | 9.529027334564862 |
| 3 | 17.86232797672583 |
| 4 | 28.79320456326416 |
| 5 | 42.32195716406112 |
| 6 | 58.44868675610014 |
| 7 | 77.17343637198603 |
| 8 | 98.49622735850399 |
| 9 | 122.4170714846681 |
| 10 | 148.9359757636747 |

**Table 2**

| N | m = 0 | m = 1 | m = 2 | m = 3 |
|---|---|---|---|---|
| 2 | 0.649491141019818 | 4.220288503393705 | 11.80493460157114 | 24.47395314201603 |
| 3 | 0.639369160661029 | 3.811775932688927 | 9.755522988492659 | 18.75894529917058 |
| 4 | 0.639314968595286 | 3.792439635980491 | 9.544808359632656 | 17.97160609977003 |
| 5 | 0.639314877261703 | 3.791973278623325 | 9.529634749795701 | 17.87094973773380 |
| 6 | 0.639314877200001 | 3.791967959126813 | 9.529040785868604 | 17.86274690292003 |
| 7 | 0.639314877199981 | 3.791967926881534 | 9.529027518799769 | 17.86234110780681 |
| 8 | 0.639314877199981 | 3.791967926767706 | 9.529027336215217 | 17.86232825577884 |
| 9 | 0.639314877199981 | 3.791967926767456 | 9.529027334574952 | 17.86232798090355 |
| 10 | 0.639314877199981 | 3.791967926767455 | 9.529027334564905 | 17.86232797677124 |
| 11 | 0.639314877199981 | 3.791967926767455 | 9.529027334564862 | 17.86232797672620 |
| 12 | 0.639314877199981 | 3.791967926767455 | 9.529027334564860 | 17.86232797672583 |

**Table 3**

| m = 0 | m = 1 | m = 2 | m = 3 |
|---|---|---|---|
| ±0.639314877199981 | ±3.791967926767456 | ±9.529027334564903 | ±17.86232797677164 |
| ±7.546955713380563 | ±14.11211461699186 | ±23.39853648473328 | ±35.33522093207102 |
| ±21.30149434844253 | ±31.30594615567633 | ±44.08929993602973 | ±59.62388713543480 |
| ±42.90258786306490 | ±57.44274812449058 | ±75.72582617029448 | ±98.15003994606802 |
| ±122.6960339108672 | ±163.1264946245198 | ±215.0441486890701 | ±279.7562914408030 |

**Table 4**

| m = 0 | m = 1 | m = 2 | m = 3 |
|---|---|---|---|
| ±0.639314877199981 | ±3.791967926767455 | ±9.529027334564862 | ±17.86232797672583 |
| ±7.546955713288350 | ±14.11211459502552 | ±23.39853510710728 | ±35.33518997949315 |
| ±21.30090330940647 | ±31.30169788881059 | ±44.06625954114674 | ±59.53087174459596 |
| ±41.92730715200201 | ±55.36544285378535 | ±71.58600932337222 | ±90.53920882971596 |
| ±69.42838570006542 | ±86.30405871261984 | ±105.9719786012650 | ±128.3935500507762 |

**FIGURE CAPTION**

**Fig. 1**: The critical dipole moment functions $\{\hat{\xi}_n(\mu)\}_{n=0}^{10}$ (in atomic units) associated with the generalized dipole polynomials defined in Appendix B. The dimension of the polynomial space $N = 100$ and $\mu \in [0, 30]$.

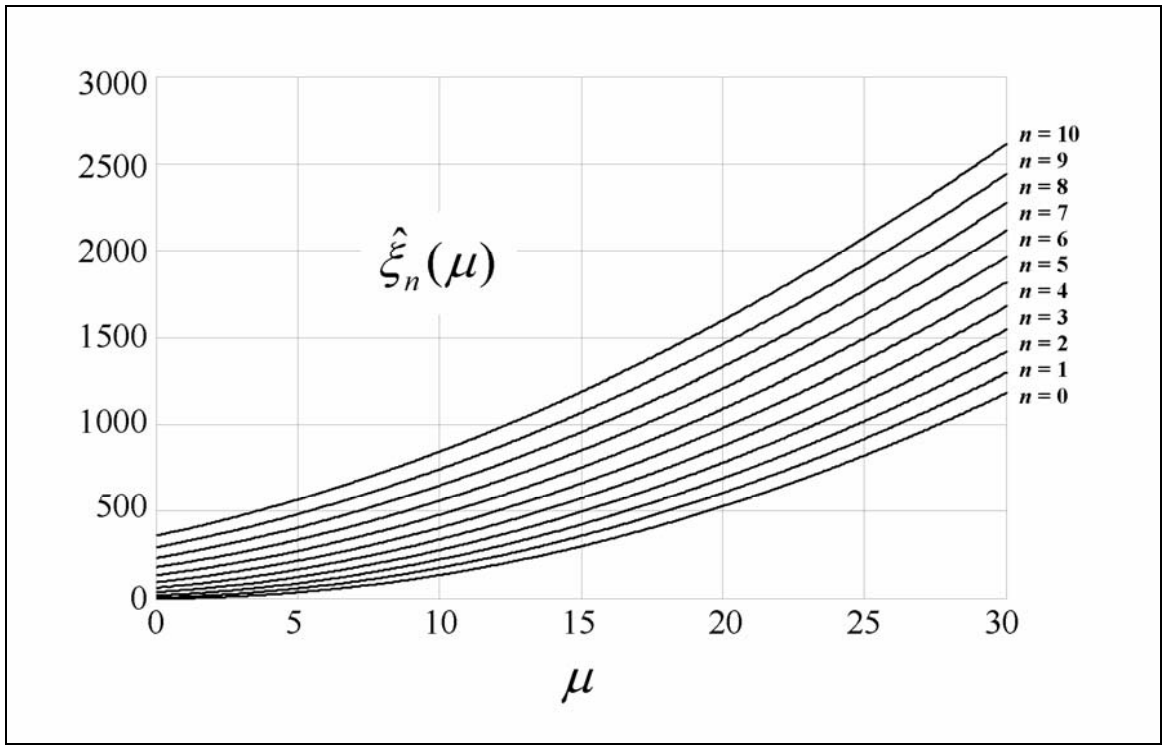

**Fig. 1**